\documentclass[]{aa}
\usepackage{natbib}
\usepackage{psfig}
\usepackage{txfonts}
\def\EBV{\mbox{E$_{\rm B-V}$}}

\def\HH{\mbox{H$_2$}}

\def\nH2{{\rm n}({\rm H}_2)}
\def\NH2{{\rm N}({\rm H}_2)}
\def\pccc{~{\rm cm}^{-3}} 
\def\pcc {~{\rm cm}^{-2}}

\def\Tstar#1 {\mbox{${\rm T}_{\rm #1}^*$}}
\def\Tsub#1 {\mbox{$T_{\rm #1}$}}
\def\TK  {\Tsub K }

\def\p{\mbox{$^+$}}

\def\hcop{\mbox{{HCO\p}}}

\def\hhco{\mbox{H$_2$CO}}
\def\h13cop{\mbox{{H$^{13}$CO\p}}}

\def\c3h2{\mbox{C$_3$H$_2$}}
 
 \def\R0{R$_0$}

\def\ddeg{{}^\circ\kern-.1em}  
 \def\pc{\rm pc}

\def\ps{\mbox{s$^{-1}$}}

\def\E#1 {$10^{#1}$}
\def\E#1 {E{#1}}
\def\P#1,{$\nH2\TK~=~#1\times~10^4\pccc$~K}
\def\ec#1,#2,#3,{#1\,(#2)\E{#3}}

\def\H3{\mbox{H$_3$}}

\def\ammon{\mbox{N\H3} }

\def\zetaH{\mbox{$\zeta_H$}}
\def\RH2{\mbox {R$_{\rm G}$}}
\def\fH2{\mbox {f$_{\HH}$}}
\def\FH2{\mbox {F$_{\HH}$}}

\title{Time-dependent \HH\ formation and protonation in diffuse clouds}
\author{H. S. Liszt\inst{1} }
\institute{National Radio Astronomy Observatory,
           520 Edgemont Road,
           Charlottesville, VA,
           USA 22903-2475}

\begin{document}
\date{received \today}
\offprints{H. S. Liszt}
\mail{hliszt@nrao.edu}
%
% \abstract{}{}{}{}{} 
% 5 {} token are mandatory

\abstract
  % context heading (optional) leave it empty if necessary  
   {}
  % aims heading (mandatory)
    {To demonstrate the time-approach to equilibrium of \HH-formation 
   and protonation in models of diffuse or H I interstellar gas 
   clouds previously published by the author.}
  % methods heading (mandatory)
   {The microscopic equations of \HH-formation and protonation are
integrated numerically over time in such a manner that the overall
 structures evolve self-consistently
under benign conditions.}
  %results heading (mandatory)
{The equilibrium \HH\ formation timescale in 
an H I cloud with N(H) $\approx 4\times10^{20}~\pcc$ is 
$1-3\times10^7$ yr, nearly independent of the assumed density or 
\HH\ formation rate on grains, {\it etc}.  Attempts to speed up
the evolution of the \HH-fraction would require densities well
beyond the range usually considered typical of diffuse gas. 
The calculations suggest that, under benign, quiescent conditions, 
\HH\ in the diffuse ISM formation of \HH\ is favored in larger 
regions having moderate density, consistent with the rather high mean 
kinetic temperatures measured in \HH, 70-80 K.

Formation of \H3\p\ is essentially complete when \HH-formation 
equilibrates but the final abundance of \H3\p\ appears more nearly 
at the very last instant.
Chemistry in a weakly-molecular gas has particular properties so
that the abundance patterns change appreciably as gas becomes more fully 
molecular, either in model sequences or with time in a single model.  
One manifestation of this is that the predicted abundance of \H3\p\ 
is much more weakly dependent on 
the cosmic-ray ionization rate when n(\HH)/n(H) $\la 0.05$.  In
general, high abundances of \H3\p\ do not enhance
the abundances of other species ({\it e.g.} \hcop) but 
late-time OH formation proceeds most vigourously in more diffuse 
regions having modest density, extinction and \HH\ fraction and somewhat 
higher fractional ionization, suggesting that atypically high OH/\HH\ 
abundance ratios might be found optically in diffuse clouds having 
modest extinction.}
 %conclusions headiing (optional), leave it empty if necessary
{}

\maketitle

\section {Introduction.}

The fact that individual H I/diffuse clouds have a substantial 
component of 
molecular hydrogen has been recognized observationally since the 
original {\it Copernicus} observations \citep{Jur74,Spi78} and
a quite high fraction of hydrogen in the nearby diffuse interstellar 
medium (ISM) overall is in molecular form \citep{SavDra+77,LisLuc02}.   
More surprising perhaps is the recent discovery that this molecular 
component hosts a rich polyatomic chemistry with relatively large 
amounts of \H3\p\ \citep{McCHin+02} and a dozen other species 
including such ions as \hcop\ \citep{LucLis96} and 
HOC\p\ \citep{LisLucBla04} and molecules as complex as \c3h2\ 
\citep{CoxGue+88,LucLis00C2H}, \hhco\ \citep{LisLuc95a,MooMar95, LisLuc+06}
and \ammon\ \citep{Nas90,LisLuc+06}.

The presence of copious \HH\ and relatively large amounts of \H3\p\
in H I clouds is supported theoretically when self-shielding 
\citep{LeeHer+96} is included in equilibrium models of diffuse gas 
\citep{LisLuc00,Lis03} using empirically-determined \HH\ formation 
rates \citep{Jur74,Spi78,GryBou+02}; observed HD/\HH\ ratios are
also explained as long as the cosmic-ray  ionization rate is taken 
large enough to support the inferred proton density in the presence 
of atomic-ion neutralization on small grains.  However, formation of 
\HH\ is a notoriously slow and supposedly rather fragile process.  
Moreover, it 
remains to be determined whether any such equilibrium is actually 
attained in the interstellar medium, which is the subject of this work.
 
Here we show the approach to equilibrium in the models discussed by 
\cite{LisLuc00}, \cite{Lis02} and \cite{Lis03}:  we present calculations 
of time-dependent \HH\ and \H3\p\ formation in small, mostly-static 
gas parcels meant to represent typical diffuse ``clouds'', like, 
for instance, the ``standard'' H I cloud of \cite{Spi78}.  As we 
discuss, the distinguishing characteristic of the chemistry in such 
clouds is that the \HH\ formation process is unsaturated 
(n(\HH) $<$ n(H)/2).  For this reason the equilibrium timescales for 
\HH\ formation cannot be calculated from simple scaling arguments 
based on the local microscopic \HH\ formation rate, nor are they 
shortened by self-shielding, or in many cases by assuming higher 
density or even a higher rate constant for \HH\ formation on grains, 
{\it etc}.  Furthermore the abundance of \H3\p\ does not always vary 
as might be expected from consideration of its equilbrium chemistry 
in more fully molecular gas.

Section 2 gives details of the methods used in the calculations, 
results of which are presented and discussed in Sect. 3 and 4.

\section{Details of the calculations and microphysics}

\subsection{Framework and initial conditions}

The calculations presented here build on a framework which was 
established in the work of \cite{WolHol+95}.  Specifically, the 
two-phase model of heating and cooling developed by \cite{WolHol+95} 
(see also \cite{WolMcK+03}) was applied to small, uniformly
illuminated spherical gas clots of constant density, intended to
represent diffuse 'clouds'; our model, however, uses the gas-phase
abundances of \cite{SavSem96} which makes them a bit hotter.  
Assumption of a geometry is required because the accumulation of 
\HH\ and CO depends on shielding by dust and other molecules and 
is, therefore, strongly non-linear.
We used the shielding factors calculated by \cite{LeeHer+96}.
Typical densities and column densities considered here, n(H) 
$= 32 \pccc$, N(H) $=4\times10^{20}\pcc$ correspond to a Spitzer 
``standard cloud'' \citep{Spi78} having a dimension of a few pc 
and median reddening 0.05 mag over the face of the cloud.  The 
central column density in such an H I cloud corresponds to the 
regime where the appearance of high \HH-fractions is first noted 
in the ISM.

%1 was here

In general, the hydrogen densities in our models (number or column) 
are assumed values.  The temperature and thermal pressure follow 
from the details of the calculation, chiefly through the
assumed gas-phase abundances of the dominant coolants carbon
and oxygen, and the influence of the radiation field in heating 
the gas {\it via} the photoelectric effect on the same small grains 
which are so largely responsible for the overall charge balance.  
As shown in Fig. 9 of \cite{LisLuc00}, a given value of the 
scaling factor (G$_0$) for the overall radiation field produces
a roughly constant thermal pressure in the outer regions
of our models, increasing somewhat with assumed density
and decreasing slightly going inward from the cloud edge.
The generally-accepted value for the thermal pressure of the
ISM, p/k $\approx 2-3\times10^3\pccc$ K  \citep{Jen02,JenTri01}
indeed applies to our typical model of the Spitzer ``standard 
cloud'' and the very densest models considered here have 
pressures a factor 2-3 times higher, as seems observationally 
to be the case as well.

 The initial condition here is simply the equilibrium diffuse cloud 
model which would be calculated in the absence of any molecular 
chemistry, as is more typical in discussions of the overall structure 
of the diffuse ISM \citep{WolMcK+03}. That is, the initial model
is equilibrated and self-consistent in terms of thermal and 
ionization equilibrium, but no molecule formation has occurred.
Our models include the rudiments of molecular chemistry as discussed 
in several recent papers concerning questions of equilibrium CO, \HH, 
HD and \H3\p\ formation in diffuse gas \citep{LisLuc00,Lis02,Lis03}.  
Incorporation of the chemistry should not normally upset the basic 
microphysical balance (ionization, thermodynamic, {\it etc.}) in 
diffuse gas because such small
fractions of oxygen and carbon are sequestered, although inclusion
of OH and \HH\ formation can affect some aspects of the distribution of
charge among species.

 Another effect mitigating the influence of \HH\ formation in diffuse 
gas is the dominance of atomic-ion neutralization by small grains, 
substantially lowering the H\p\ and electron density which would 
otherwise obtain in the presence of cosmic-ray ionization of 
hydrogen \citep{WolHol+95,Lis03}.  This effect is of the utmost 
importance in understanding the interplay between the cosmic ray 
ionization rate \zetaH\ (here taken as the rate per free H-nucleus) 
and abundance of \H3\p\ \citep{Lis03} .  Detailed calculation 
including grain charging \citep{DraSut87,WeiDra01Supp} is required 
if the heating rate is to be properly calculated according to the 
prescription of \cite{BakTie94}.

\subsection{Formation of \HH}

In this work we distinguish between the {\it formation} of \HH ,
which we consider to be a microscopic physical effect occuring
on individual large grains and involving individual or paired 
H-atoms, and the {\it accumulation} of \HH\ in the ISM, which
is a macroscopic effect involving assumptions about cloud geometry, 
ISM topology, and so forth.  The purpose of this work is to understand 
the macroscopic aspects of the accumulation of \HH\ in the ISM
and we  employ only the most standard, empirical, microscopic 
description of its formation.

That is, the volume formation rate of \HH\ is expressed as dn(\HH)/dt 
= n(H) n(H I) \RH2\ $<$v$_T$$> = 3\times 10^{-18} $cm$^3$ \ps\ 
n(H) n(H I) $\sqrt{\TK}$ \citep{Spi78} where the factor n(H) 
(the total density of H-nuclei)  represents a constant number of large
\HH-forming grains per H-nucleus in the gas; n(H I) is the density
of atomic hydrogen (after protons, H$^-$ ions, and \HH, {\it etc.} 
are reckoned) and the thermal velocity and/or kinetic temperature 
terms correspond to the speed at which
H-atoms move through the gas, possibly encountering large grains.
The rate constant \RH2\ nominally includes a sticking coefficient.

The inferred rate in the diffuse ISM has been remarkably steady over 
the last 25 years \citep{Jur74,Spi78,GryBou+02} and use of this simple
formulation, combined with the modern shielding coefficients, accounts 
extremely well for the observations of \HH\ in diffuse lines of sight 
\citep{LisLuc00}.  As discussed in Sect. 3 (see Fig. 4) varying the
 assumed value of \RH2\ by factors of a few has appreciable effect
upon the amount of \HH\ which is produced but the equilibrium time
scale changes little.
  As an aside we note that the gas-phase formation of 
\HH\ {\it via} H$^-$, which could account for some of the very low 
molecular fractions seen along lines of sight having low extinction 
\citep{Lis02}, is included in the present calculations as well but 
without much effect. 

\subsection{Solving the \HH-accumulation problem}

The equilibrium solutions for the radial distributions of CO, \HH\
and \H3\p\ described in our earlier papers were achieved through what
is essentially a relaxation method.  An initially atomic
sphere is divided into a substantial number (typically 100) of thin 
radial shells and the molecular densities are calculated for each, working 
inward, self-consistently accounting for the shielding which accrues
from each configuration.  Each time a quasi-equilibrium state is
reached for the entire structure (that is, when the calculation has
successively converged in each shell), the calculation starts anew at 
the outermost shell (whose \HH-abundance was, after all, calculated in 
at least partial ignorance of the conditions interior to it) and the 
process is repeated until the whole structure varies sufficiently little 
between iterations.  The process is not particularly difficult or 
unstable and convergence is rapid, typically requiring only a few 
seconds on a recent laptop computer.  Note that the thermal and charge
balance, {\it etc.} are continually updated during the calculation.

The time-dependent calculations basically just wrap the equilibrium 
solver in an outer loop which allows the entire structure to evolve 
self-consistently over short times.  The time step, which initially 
cannot be as large
as even 1000 years in most cases,  is examined and lengthened by a 
factor of $\sqrt{2}$ whenever the structure changes sufficiently little 
between time steps.  The full time-dependent calculation {\it is} somewhat 
tedious, typically requiring several hours per model and much exercising
of the tiny (but assertive) fan on the author's laptop computer in cases
of high \HH-fraction.

\subsection{Elementary analytic considerations of the time-dependent
solutions for \HH}

Although the real solution to the time-dependent accumulation problem 
is strongly non-linear and can only be achieved 
globally over an entire geometrical construct, several important
aspects of the problem may be understood with reference to  a simple 
parametric treatment of purely local behaviour.

Accordingly, the local rate of accumulation of molecular hydrogen
may be written 

$$ dn(\HH)/dt = -n(\HH)\Gamma_{\HH} + \RH2 n(H) n(H I) \sqrt{\TK} \eqno(1) $$

where $\Gamma_{\HH}$ is the rate of \HH\ destruction (mostly by 
cosmic rays and ultra-violet photons, although interaction of 
He\p\ and \HH\ is included in the numerical calculation), \TK\ is the 
kinetic temperature (calculated in the models) and \RH2\ 
$= 3\times 10^{-18}$ cm$^3$ \ps\ (see Sect. 2.2).  After rewriting 
n(H I) = n(H) - 2 n(\HH) (for the present purposes;
all forms of hydrogen are considered in the numerical models) the 
solution is 

$$n(\HH)_{|t} = b/a + (n(\HH)_{|t=0} - b/a) e^{-at} \eqno(2)  $$

where $a = \Gamma_{\HH} + 2 \RH2 n(H) \sqrt{\TK}, 
b = \RH2 n(H)^2 \sqrt{\TK}$.

%1
\begin{figure*}
\psfig{figure=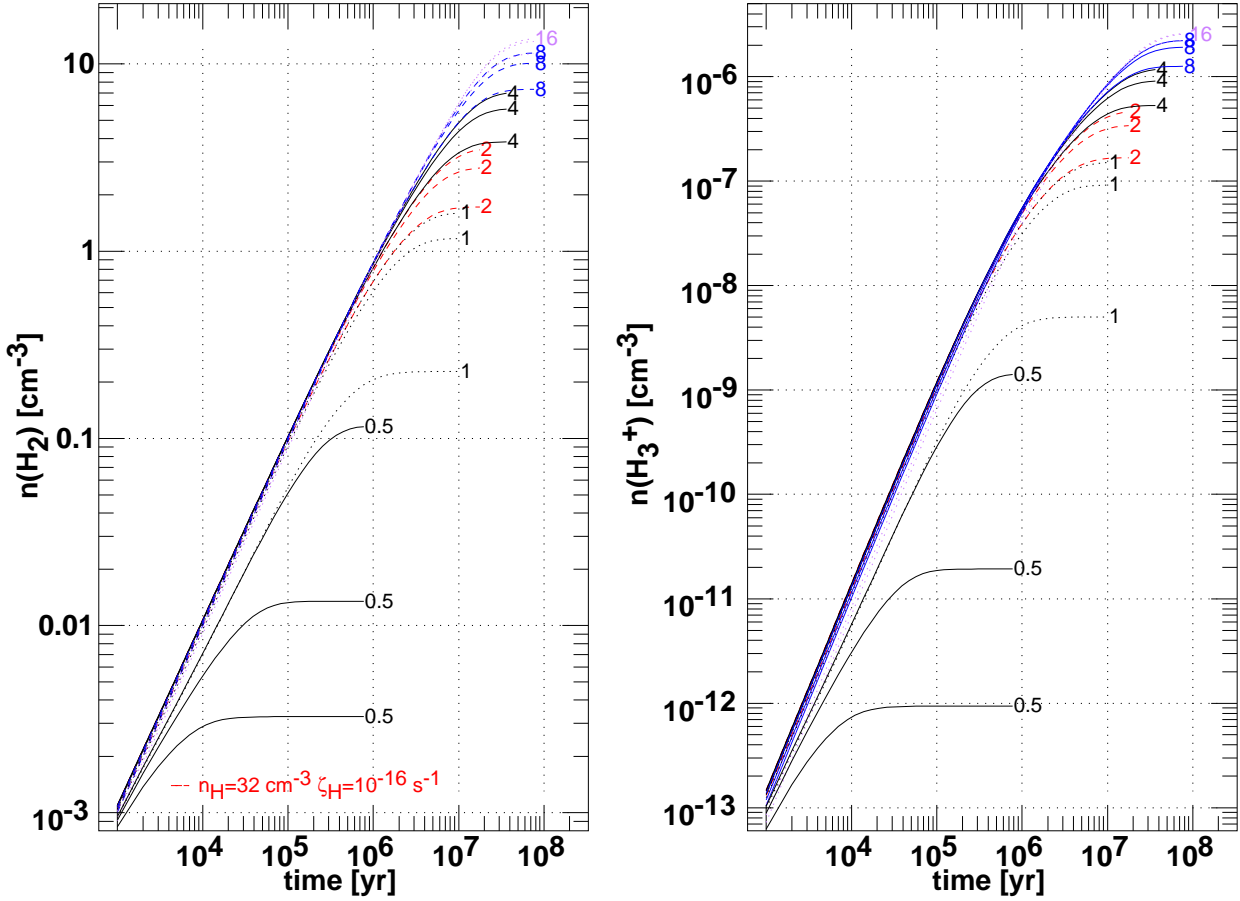,height=12cm}
\caption[]{Density of molecular hydrogen n(\HH) (left) and \H3\p\ (right) 
as functions of time within spherical, uniformly illuminated, initially
atomic gas spheres of constant density of H-nuclei n(H) $=32\pccc$,
threaded by a cosmic ray flux yielding a primary ionization rate per 
H-atom \zetaH\ $= 10^{-16}$ \ps.  Three curves corresponding to shells 
at fractional radii 0, 1/3 and 2/3 from the center are shown for each of 
a series of models with differing central column density of hydrogen
$0.5 \times 10^{20}\pcc \leq $ N(H) $\leq 16 \times 10^{20}\pcc$.
The physical size of the models in pc and their central (edge-to-edge)
column density in units of $10^{20}~\pcc$ are equal and shown at the
right of each curve.}
\end{figure*}

The time constant in the problem is 1/a, where a is a linear 
combination of the per-particle formation and destruction rates, 
observing the conservation of H-nuclei,  and is therefore 
relatively short in weakly-shielded gas.  The rate of accumulation 
of \HH\ takes on its highest value dn(\HH)/dt = b at time 0 and 
the equilibrium solution n(\HH)$_{|t \rightarrow \infty}$ = b/a 
may be thought of as the 
result of accumulation at the maximum rate over a period of time 
1/a.  Thus, self-shielding (lowering $\Gamma_{\HH}$) cannot increase 
the rate at which \HH\ forms  or speed up the accumulation of 
\HH, it only increases the final amount of \HH\ by allowing the 
accumulation process to work longer.  By lengthening the time-scale,
self-shielding may also provide a buffer against sudden change once
equilibrium is attained.

%2 was here

Increasing the microscopic formation rate constant or density 
will hasten the pace at which \HH\ accumulates but even so,
global equilibria are not generally achieved more rapidly in cases 
where the \HH\ fraction would otherwise be small, as discussed in
Sect. 3 (see Fig. 4); it only happens that more \HH\ is made (which 
requires somewhat more time).  Maintenance of high equilibrium 
\HH-fractions requires 
$\Gamma_{\HH} \la \RH2 n(H) \sqrt{\TK}$, which is roughly
equivalent to reducing the photodestruction rate (of order 
$10^{-10}$ \ps\ in free space) to insignificance {\it i.e.}
$\Gamma_{\HH} \approx \zetaH $.

\subsection{Weak {\it vs.} slow processes}

Indeed, the situations which equilibrate most rapidly are those for 
which the \HH\ abundance is smallest, which exemplifies the fact that 
even processes like \HH\ formation or cosmic-ray ionization -- which 
seem weak because they have low rates -- may equilibrate quickly when 
not called upon to do very much; the time constant is then determined
by the overall destruction rate, which may be high.  It is for this 
reason that formation of \H3\p\ -- a much higher-order process which 
after all requires formation of \HH, cosmic ray ionization of \HH\ and 
further interaction of \HH\p\ and \HH\ (see Sect. 4.2) -- is 
essentially complete as soon as \HH\ accumulation equilibrates: one 
need not wait for a further period 1/\zetaH.

\subsection{The cosmic-ray ionization rate \zetaH}

%3 was here

The evolution of the \HH\ fraction is largely independent of
the assumed value of \zetaH\ in our models but
the standard cosmic-ray ionization rate assumed in the models in this 
work is $\zetaH\ = 10^{-16}$ \ps (the cosmic-ray destruction rate per 
\HH\ is slightly more than 2\zetaH).  This is approximately one order 
of magnitude larger than what is often used in calculations of 
cosmic-ray ionization of the ISM \citep{WolHol+95} especially for 
dense dark cloud chemistry and is about that which
is needed to explain the overall abundance of \H3\p\ (see Sect. 4).  
The equilibrium \H3\p\ concentration is nearly proportional to
\zetaH\ in our models having larger \HH\ fractions (see Fig. 5) 
owing to the intervention of small-grain neutralization of protons:  
the formation can be pushed harder without unduly increasing the 
electron density and recombination rate for \H3\p.   However, for 
reasons discussed in Sect. 4, the concentration of \H3\p\ is more 
weakly dependent on \zetaH\ in regions of small \HH-fraction or when
\zetaH\ is very large.

\subsection{ Heating by cosmic rays and \HH-formation}

The current assumption is that the diffuse ISM is heated by
the photoelectric effect on small grains in both the warm phase
(where x-rays make a larger contribution) and cool phase.   
The heating rate as a function of density in partly-shielded 
gas is shown in Fig. 3 of \cite{WolHol+95} based on the work 
of \cite{BakTie94}.  As may be seen from that Figure, the cosmic 
ray heating rate is roughly 1.2 dex below that of the photoelectric
effect in diffuse gas and 2-2.5 dex lower in cool gas, for 
\zetaH = $10^{-17}$\ps.  Thus the higher cosmic-ray ionization 
rates considered here may imply a substantial contribution to 
the heating of the warm gas by cosmic rays, harkening back to 
the very earliest discussions of two-phase equilibrium, before 
the role of diffuse X-ray heating was incorporated \citep{GolHab+69}. 
The highest cosmic-ray rates considered here still do not
produce dominant heating effects in the cool neutral gas,
but much higher ionization rates $\zetaH\ \ga 10^{-15}\ps$ 
were  recently suggested by \cite{LePRou+04} for the line of 
sight to $\zeta$ Persei.

 In our models there is a modest contribution to the heating due to 
\HH\ formation, comparable to that arising from heating by cosmic-ray 
ionization of H I.  The volume heating rate due to the latter is 
$\zetaH$ n(H I) $\langle{\rm E}_{cr}\rangle$ where  
$\langle{\rm E}_{cr}\rangle \approx 8 $eV is the energy available for heating 
after the primary ionization.  The equivalent rate for \HH-formation is
$\RH2 n(H) n(H I) \sqrt{\TK} \langle{\rm E}_{\HH}\rangle$ where 
$\langle{\rm E}_{\HH}\rangle$ 
represents that fraction of the heat of formation of \HH\ which is 
available to heat the gas.  The heating rates for the two processes
are very nearly equal under typical conditions, 
$\zetaH = 10^{-16}$ \ps, 
$\TK = 80$ K, n(H) $= 32 \pccc$ and $\langle{\rm E}_{\HH}\rangle$ = 1.5 eV.

\subsection{The recombination rate of \H3\p\ and other
chemical reactions}

The chemical rate constants used in our models have been taken
from the UMIST database 
\citep{LeTMil+00} (see {\it http://www.rate99.co.uk})
including, for consistency with the results of \cite{Lis03}, 
the rate constant for e + \H3\p\ recombination 
$\alpha(\TK) = 4\times10^{-6}/\sqrt{\TK}$ cm$^3$ \ps.
This rate has been measured many times, including, most recently
and probably accurately, by \cite{McCHun+03}.  Their measured
values, at 23 K and 300 K, are 70\% of those used in our work.

\section{Time-dependent \HH\ accumulation in diffuse clouds}

\subsection{Evolutionary effect of varying column density}

Figure 1 shows the evolution of the \HH\ and \H3\p\ densities in a 
series of models, each having the same uniform total density of 
hydrogen n(H) $= 32\pccc$ threaded by a cosmic ray flux yielding
$\zetaH\ = 10^{-16}~\ps$.  The central column density of hydrogen
across the clouds N(H) (from edge to edge) varies from 0.5 to 16 $\times
10^{20}\pcc$, as indicated at the right hand side of each curve. 
The physical size of the models (diameter) in pc is, coincidentally, 
equal to the  column density as labelled in units of $10^{20}~\pcc$.  
Three curves are shown for each model, corresponding to the shells 
centered at fractional radii 0, 1/3 and 2/3 from the center.  These 
curves separate at larger times when the outermost regions have 
reached equililbrium.  For a sphere,
the median line of sight occurs at a fractional impact parameter of
2/3 where the column density traversed is 3/4 that at the center.
The median line of sight through the model with central column
density $4\times10^{20}\pcc, 3\times10^{20}\pcc$ corresponds
exactly to the standard H I cloud of \cite{Spi78}.

For the optically-thinnest model there is a large radial gradient 
in the \HH\ density and a wide disparity between the times required 
to reach equilibrium ($10^4 - 10^6$ yr).  The fact that such appreciable
(easily detectable) amounts of \HH\ form in as little as 10$^4$ yr
demonstrates why it is so hard to find lines of sight having 
measurable reddening which are devoid of \HH .  For N(H) as large as
$16\times10^{20}\pcc$ the \HH\ accumulation process is thoroughly 
saturated (n(\HH) $\rightarrow$ n(H)/2) when equilibrium is reached 
at very large times approaching $10^8$ yr.  Models in which the 
equilibrium \HH-distribution is more uniform also evolve more uniformly, 
as illustrated in Fig. 2  which shows (at top) the time-dependent 
evolution of the radial variation of n(\HH) in the model resembling
a ``standard'' H I cloud.  The \HH-density is remarkably constant
for fractional radii of 90\% or more for most of the first 
$1-3\times 10^{6}$ yr, after which a more molecular core develops, 
as well as a very thin (actually, unresolved) transition layer 
into a more nearly atomic envelope.  

Clearly, self-shielding is important only at somewhat later times.
In some models with moderate density, n(\HH) behaves as a single 
power-law from center to edge, as illustrated in Fig. 2 at bottom 
which shows the equilibrium solutions for n(\HH) $=32 \pccc$
corresponding to the models used in Fig. 1.  However this is not a
general feature of the equilibrium solutions.   Figure 2 also
illustrates the effect of finite resolution on the structure
of the atomic envelope, which fractionally is substantially
larger with higher resolution.  The results in Fig. 1 are not
affected by the illustrated doubling of the number of radial
shells.

%2
\begin{figure}
\psfig{figure=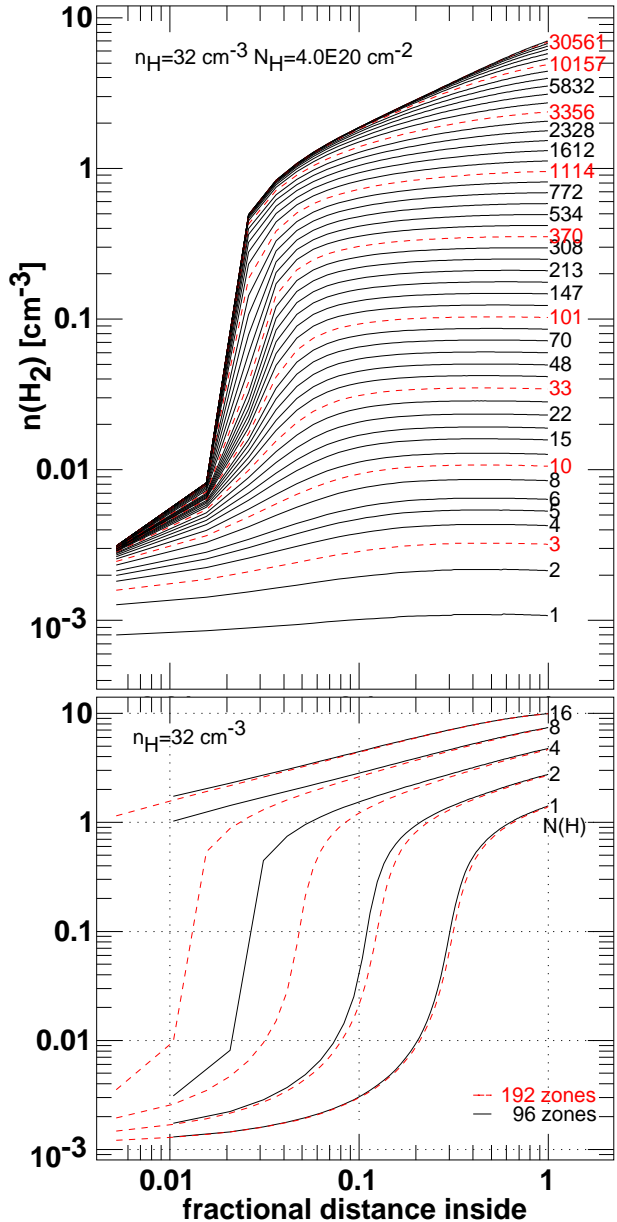,height=16.3cm}
\caption[]{Radial and secular variation of molecular hydrogen 
density within models illustrated in Fig. 1.  Top: Time evolution 
of n(\HH) for n(H) = 32 $\pccc$, N(H) = $4 \times 10^{20}\pcc$.  
Labels at right give the elapsed time in units of 1000 yr.  Bottom:
Radial variation in equilibrium for n(H) = 32 $\pccc$ and
N(H) $= 1,2,4,8,16 \times 10^{20}~\pcc$, for calculations employing
96 and 192 radial zones.} 
\end{figure}

As shown in Fig. 1, even a standard H I cloud model has a 
very substantial fraction of molecular hydrogen in its interior.
This of course is borne out observationally \citep{Jur74} as 
molecular fractions of 25\%-45\% are derived for even the diffuse 
ISM observed in {\it uv}/optical
absorption lines of \HH\ and CH \citep{SavDra+77,LisLuc02}.
Less obvious, however, is that the conditions so favourable to 
\HH\ may actually work for as long as the calculations require to 
reach equilibrium  in Fig. 1, {\it i.e.} for times as long as 
$10^8$ yr.  

\subsection{Evolutionary effect of varying number density}

It seems natural to inquire in what ways the evolution might be 
hastened.  Shown in Fig. 3 are results for series of models 
of different number density $8\pccc \le $ n(H) $\le 128\pccc$ at 
column densities N(H) = $2-16 \times 10^{20} \pcc$.
At least for moderate column densities, as shown at the left in Fig. 
3, increasing the number density within the range expected for 
diffuse gas has surprisingly little effect on the time required for 
the models to reach equilibrium: the increased rate of accumulation 
is compensated by the tendency to produce more \HH , making the 
equilibrium timescale about constant over a wide range of density 
for the standard H I cloud column density.  

Only when very high \HH\ fractions are attained at a given density
does increasing the density shorten the equilibrium time ({\it i.e.}
Fig. 3 at right, or for the two very highest densities at left).  With
reference to Fig. 3, note that, although the local accumulation rate 
initially varies as n(H)$^2\sqrt{\TK} \propto$ n(H)$^{3/2}$ in thermal 
pressure equilibrium, the total amount of \HH\ produced within a
model at any time is much less dependent on density because the 
more tenuous models are larger, varying as 1/n(H).  
Obviously,  all models in which the \HH\ accumulation saturates 
at high molecular fractions produce the same N(\HH) if they have the 
same N(H), so  less dense models having the same N(\HH) then make 
more \HH\ molecules overall because they are larger.

\subsection{Does size matter?}

Physically larger clouds tend to produce more \HH\ in both series
of models just discussed.  Those in Fig. 1 with larger column 
density N(H) are physically larger and have higher internal 
molecular number density, so, at fixed number density, physically 
larger regions of higher column density are responsible for the very 
great bulk of the accumulation of \HH.  In Fig. 3, molecular number 
densities are higher in clouds 
of higher n(H), but usually not by such large amounts as to 
compensate for the very large differences in volume.  That is, the 
model volumes vary by a factor $(128/8)^3 \approx 4000$ whereas 
n(\HH) varies by factors of 100-500 for the models with 
N(H) = $4-16\times 10^{20}~\pcc$.  Over a substantial range in
N(H), it is expected that the bulk of the molecules will form
in (again) physically larger regions,  but of lower density.

%3
\begin{figure*}
\psfig{figure=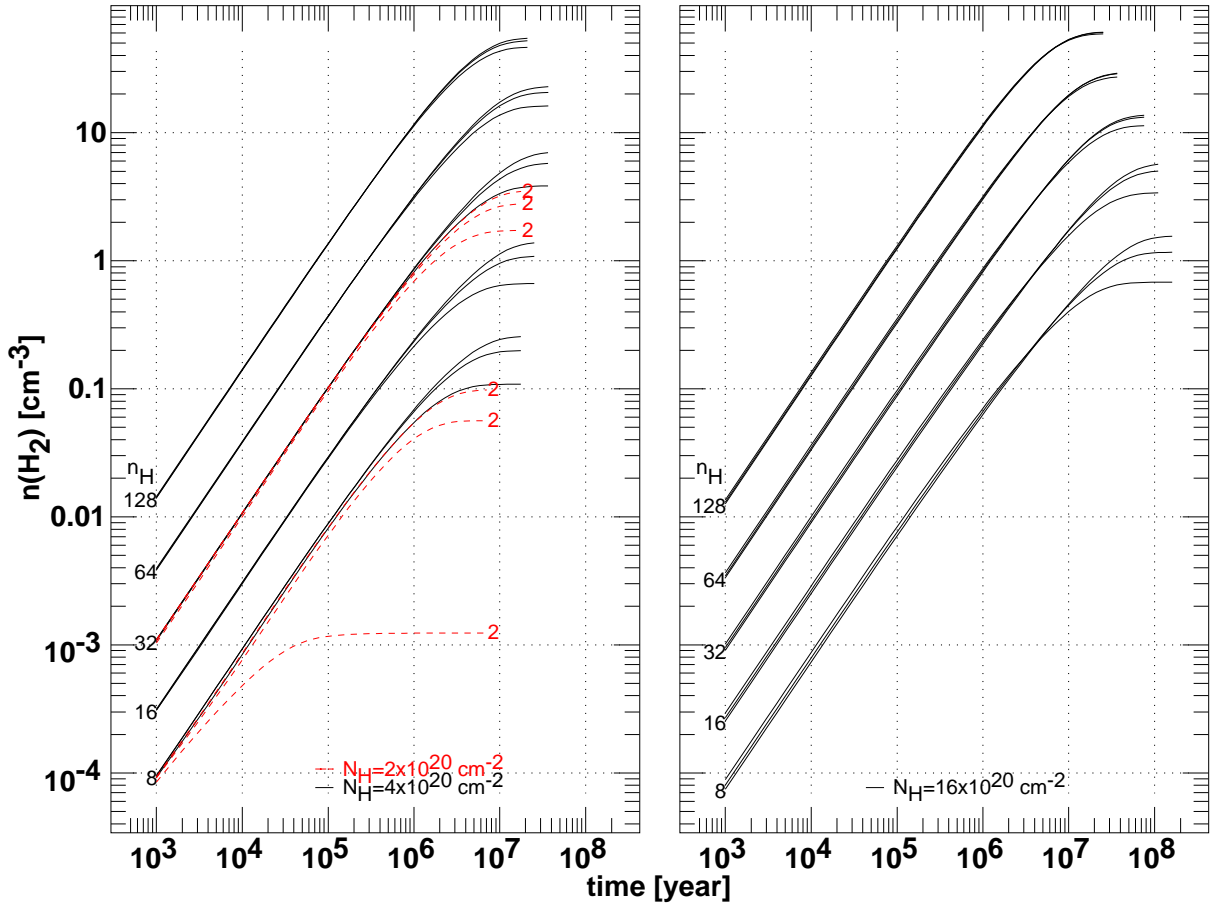,height=12cm}
\caption[]{Density of molecular hydrogen n(\HH) as a function of time 
within spherical, uniformly illuminated, initially atomic gas spheres 
of differing density of H-nuclei n(H) $=8-128~\pccc$ (as shown at the 
left of each family of curves), threaded by a cosmic ray flux yielding 
a primary ionization rate per H-atom \zetaH\ $= 10^{-16}$ \ps.  At 
right, all models have a central column density N(H) 
$ = 16 \times 10^{20}\pcc$ (from edge to edge); at left, results are 
interspersed for N(H)  = 2 (dashed; red in appropriate media)
and 4 $\times 10^{20}\pcc$.  As in 
Fig. 1, three curves are shown for each model, corresponding to 
fractional radii 0, 1/3 and 2/3 from the cloud center.}
\end{figure*}

So is the bulk of the interstellar \HH\ really formed in large tenuous 
regions, or in more compact, denser ones?  Observationally, the mean 
column density-weighted kinetic temperature inferred for \HH-forming 
diffuse gas is 70--80 K
\citep{SavDra+77,ShuAnd+04}, which, for typical interstellar thermal 
pressures, implies densities like those of the standard H I cloud
\footnote{Note that the surveys from which these mean temperatures 
are derived have somewhat low sample mean \HH-fractions, 0.17-0.18.}.  
Attempts to form the observed \HH\ more rapidly in
regions of high density must account for these high kinetic 
temperatures; perhaps it is relevant that \cite{JenTri01} found
a small but pervasive fraction of material at much higher than normal
thermal pressure in typical diffuse clouds probed in the
fine-structure excitation of C I.  This high-pressure component
is generally ignored in calculation of the mean thermal pressure
in the ISM, but would approximately double it.

\subsection{Changes in formation and dissociation rate constants}

The empirically-inferred microscopic formation rate of molecular 
hydrogen formation has changed remarkably little since \HH\ was 
first observed in the ISM \citep{Jur74,Spi78,GryBou+02}.
Figure 4 shows the effects of changing the assumed \HH\
formation rate constant \RH2\ in some of the models  illustrated 
in Fig. 1.  Although the \HH\ column and number densities 
increase faster than linearly with changes in \RH2\ at
late times, the timescale may only be shortened after 
the molecular fraction approaches its maximum possible value.

As noted in Sect. 2.4, high molecular fractions may only accrue 
in cases where the free-space photodissociation rate of \HH\ is 
attenuated (by dust and self-shielding) to the point where it is 
no more than competitive with cosmic-ray induced destruction rates.  
It follows that natural variations in the ambient interstellar 
radiation field (ISRF) may encourage or discourage \HH\ accumulation, 
but only modestly and at late times when equilibrium is approached.  
Shown in Fig. 5 at left are results for a lower-density model 
(n(H) $= 8 \pccc$) at 
N(H) $ = 4\times10^{20}\pcc$, with the ISRF lowered by factors of
2 and 4.  Clearly the \HH\ accumulation process may be enhanced
by making the material sufficiently dark, but only at the expense 
of incurring very long equilibrium times.  Note that the darker 
models actually form and accumulate somewhat less \HH\ at smaller 
times because they are substantially cooler; lower photoelectic 
heating rates accompany lower photoionization rates for molecules. 
Clouds in very locally darker regions would have to be overpressured
in order to compensate for this sluggishness, but in the ISM a
tendency to higher thermal pressure at higher density seems a real 
effect.

Figure 5 at right shows the effect of varying the primary
cosmic ray ionization rate by factors of three about the
standard value \zetaH\ $= 10^{-16}$~\ps.  Such variations have
little effect on the equilbrium \HH\ densities in standard
H I clouds as long as they are not so large as to upset the
thermal balance (which can occur for \zetaH $\ga 10^{-15}~\pc$).
The more complicated behaviour of \H3\p\ is discussed in Sect. 4.

%4 was here followed by 5

\section{\H3\p\ and other trace species}

\subsection{Evolution of n(\H3\p)}

In the ISM there is a general increase in N(\H3\p) with reddening 
corresponding to 
N(\H3\p) $= 8\times10^{14}~\pcc$\ over 6 magnitudes, or to 
N(\H3\p)/N(H) $\approx 2.3 \times 10^{-8}$ for the standard 
gas-reddening 
ratio N(H)/\EBV\ $ = 5.8\times10^{21}~\pcc$\ mag$^{-1}$
(see Fig. 14 of \cite{McCHin+02}).

As discussed previously \citep{Lis03} or as shown in the figures 
here, n(\H3\p) concentrations comparable to those required can be 
produced in equilibrium models of diffuse clouds of moderate number 
and column density for cosmic ray ionization rates 
\zetaH\ at or somewhat above $10^{-16}$ \ps\ (see Fig. 5 at right).  
Such models of \H3\p\ in diffuse gas have the added virtue that
the HD/\HH\ ratios observed in diffuse clouds -- which are very 
sensitive to the actual proton density --  are also reproduced.

As shown in the figures, the \H3\p\ fraction is sensitive to the 
same effects which increase the concentration of \HH, and in some 
circumstances to the cosmic ray ionization rate as well (Fig. 5 here).
There are some more subtle effects which are less obvious in the
figures:  in models which achieve large \HH-fractions, the electron 
density is lower by a factor of (typically) about two at late
times  because the gas cannot sustain high proton densities when 
most of the hydrogen is molecular.

Although the \H3\p\ fraction equilibrates on very much the same final
timescale as that of \HH , the problem of timescales is somewhat more
critical for \H3\p\ because its time derivative is much steeper;
the final concentration occurs more nearly at only the final
moments of the calculation.  Conversely, the \H3\p\ fraction
would deteriorate more rapidly should conditions suddenly 
become less hospitable.  Of course this effect is mitigated
slightly when the \H3\p\ fraction is calculated with respect to
\HH\ (instead of n(H)) but in most cases the lines of sight with 
known N(\H3\p) are not accessible to direct measurement of N(\HH) 
or even N(H I).  Note that \H3\p\ is also more centrally
concentrated than \HH, but the effect is really strong only in 
models within which there are larger radial gradients in n(\HH).

%4
\begin{figure*}
\psfig{figure=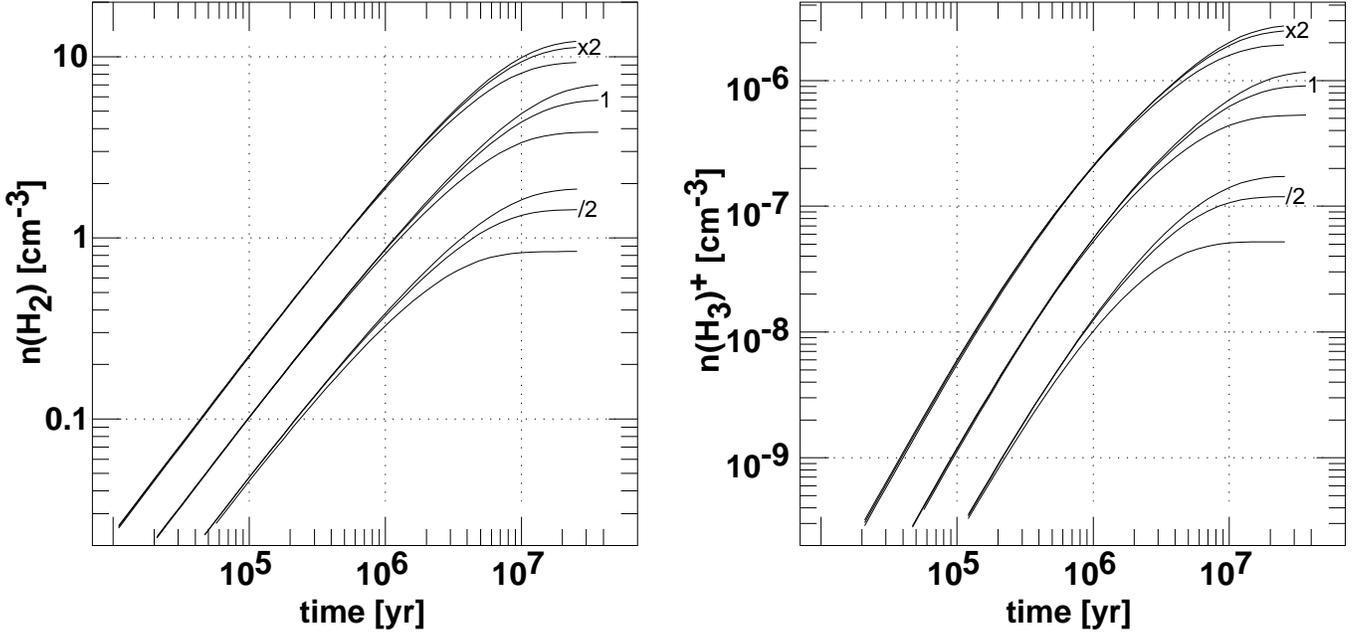,height=8.5cm}
\caption[]{Similar to Fig. 1, for models having n(H) 
$= 32 \pccc$, N(H) $=4\times10^{20}\pcc$ and primary cosmic ray 
ionization rate \zetaH\ $= 10^{-16}$ \ps , with the rate constant 
for \HH\ formation scaled by factors of 2 about the standard value.}
\end{figure*}

\subsection{Effect of changing \zetaH}

The presence of high cosmic ray ionization rates in diffuse gas
has been inferred from very elementary analysis of the \H3\p\
abundance which treats it like an atomic ion \citep{McCHin+02}.  
Assuming -- as is true of dark, dense molecular gas -- that 
every primary ionization of \HH\ leads to an 
\H3\p\ ion and that \H3\p\ ions are destroyed by electron 
recombination with a rate constant $\alpha_e$, there follows a 
simple, formal expression for the \H3\p\ density:

$$ n(\H3\p) \approx 2 \zetaH n(\HH)/(\alpha_e ~n(e)) \eqno(3) $$

The discrepancy between observation and the values of n(\H3\p)
which result from putting typical values for \zetaH, n(e), 
{\it etc.} in this expression is sufficient to suggest that
\zetaH\ $>> 10^{-17}$ \ps.  Nonetheless, Eq. 3 is only a formal, 
very approximate solution for n(\H3\p) 
because of the implicit dependences of n(e) and $\alpha_e$ upon the 
physical effects of ionization and heating (respectively) denoted 
by \zetaH, and because of other, less obvious assumptions made
about the molecular chemistry.  Furthermore, the obvious changes
in the \H3\p/\HH\ ratio with time show that such considerations
might at best apply only at very late times ({\it e.g} Fig. 5
at right).

Figure 5 at right exhibits the further surprising result that the 
\H3\p\ abundance is largely independent of the cosmic ray ionization 
rate at early times when the \HH\ abundance is small.  This occurs 
because formation of \H3\p\ is a multistep process with a resultant 
inverse dependence on the electron density which may be much faster 
than linear (as suggested by Eq. 3), combined with the tendency
(suppressed but not eliminated by neutralization of atomic ions on
small grains) for the ionization fraction to increase with \zetaH.  

In a somewhat simplified form neglecting chemical reactions 
of \HH\p\ and \H3\p\ with neutral atoms {\it etc.} (appropriate in 
diffuse gas), and the complexities of grain neutralization, the main 
steps involved in forming and destroying \H3\p\ and its immediate 
antecedent \HH\p\ are:

$$ {\rm cosmic~ray} + \HH \rightarrow \HH\p + e, \eqno(4a) $$

$$ \HH\p + e \rightarrow {\rm 2 H} , \eqno(4b) $$

$$ \HH\p + \HH \rightarrow \H3\p + {\rm H} , \eqno(4c) $$

$$ \H3\p + e \rightarrow {\rm 3 H},~etc \eqno(4d)  $$

So, whenever the recombination described in reaction 4b dominates over 
the protonation in reaction 4c, the abundance of \HH\p\ will be 
inversely  proportional to n(e) and 
n(\H3\p) $\propto$  n(\HH) n(\HH\p)/n(e) $\propto \zetaH$/n(e)$^2$.  
Calculating the ionization balance in an atomic gas according to the 
prescription in \cite{Spi78} indeed yields n(e) $\propto \sqrt{\zetaH}$
for \zetaH $\ge 10^{-17}$ \ps\ under conditions typical of H I clouds.

For the model of a typical H I cloud shown at right in Fig. 5, such 
considerations are apparently sufficient to remove much of the 
dependence on \zetaH\ from the \H3\p\ density at early times.  To 
ascertain the conditions under which this effect is important, 
note that the recombination coefficient of \HH\p\ is 
$\alpha = 4\times 10^{-6}/\sqrt{\TK}~{\rm cm}^3$ \ps\ 
and reaction 4c proceeds with a rate constant 
$k = 2.08\times 10^{-9}~{\rm cm}^3$ \ps\ in the UMIST database, 
see \cite{LeTMil+00} or {\it http://www.rate99.co.uk}.  
Thus for \TK\ = 100 K and taking n(e) as a small multiple  ($2 \times$) 
of the gas-phase abundance
of carbon, n(C\p) $\approx 1.4 \times 10^{-4}$ n(H), it follows that 
recombination dominates and the functional dependence of n(\H3\p) upon 
\zetaH\ is weakened for n(\HH)/n(H) $\la$ 0.05, with higher
\HH\ concentrations required for higher n(e) and/or \zetaH.  In Fig. 5
at right, this is case for the first $1-2 \times 10^6$ yr.

\subsection{Influence of \H3\p\ on other trace molecule
abundances}

%5
\begin{figure*}
\psfig{figure=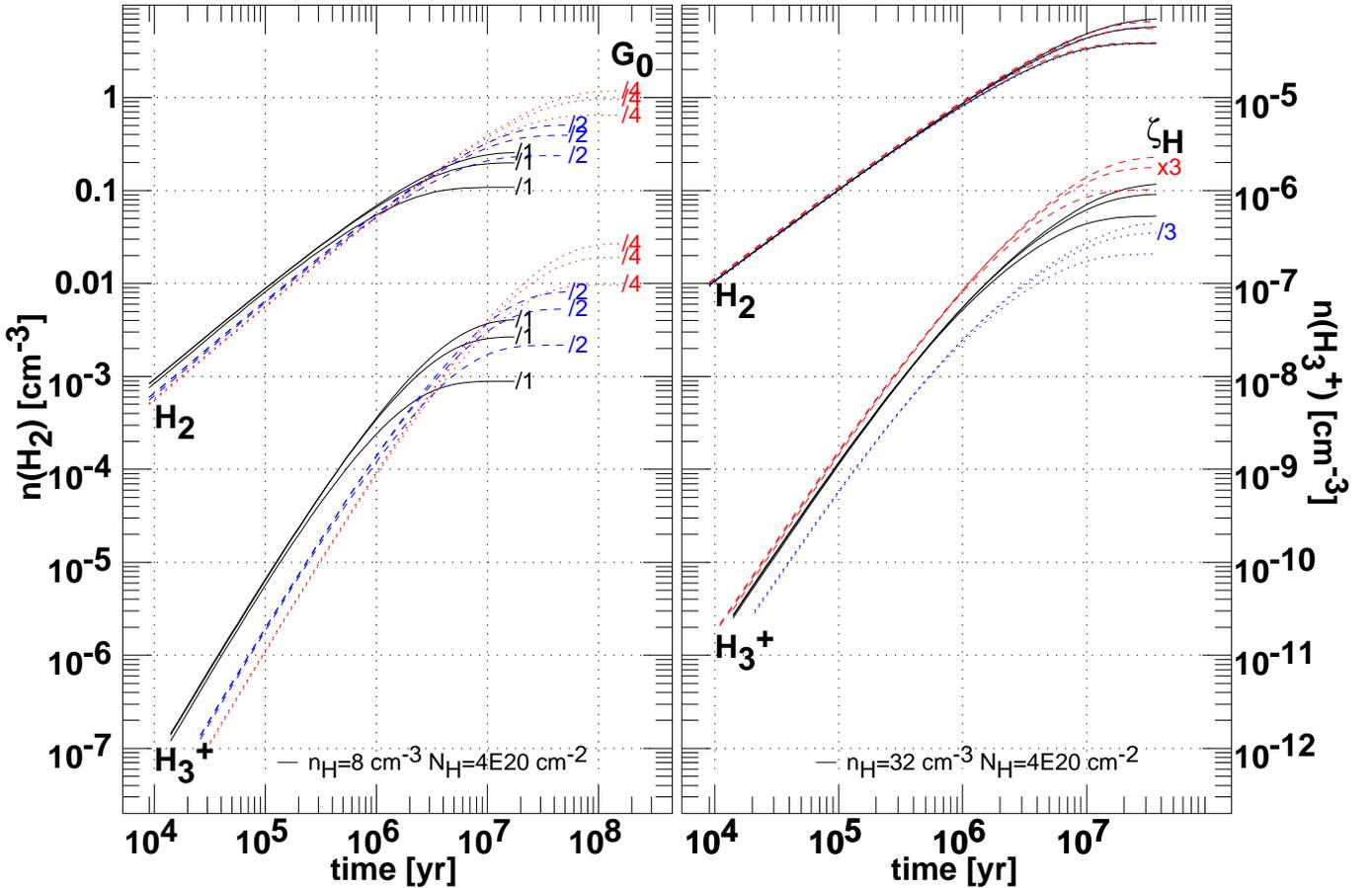,height=12cm}
\caption[]{\HH\ and \H3\p\ densities calculated for varying 
strengths of the interstellar radiation field (left) and
cosmic ray primary ionization rate \zetaH\ (right): the scale 
for n(\HH) is given at far left, that for n(\H3\p) at far right.
In the panel at left the scaling parameter G$_0$ is taken as 
1, 1/2 and 1/4; at right \zetaH\ varies by factors of 3, 
1 and 1/3 about the standard model value 
\zetaH\ $= 10^{-16}$ \ps.  As in Fig. 1, three curves are
shown for each model, at fractional radii 0, 1/3, and 2/3 
from the center.  The models in both panels have
N(H) $=4\times10^{20}\pcc$ but the model at left has lower
number density n(H) $= 8 \pccc$.  Note that n(\H3\p) is 
relatively insensitive to changes in \zetaH\ at early times 
when n(\HH) is small, as discussed in Sect. 4 of the text.}
\end{figure*}

In general the surprisingly high abundance of \H3\p\ in diffuse gas 
should not alter significantly the patterns of abundance of other 
trace species under the quiescent conditions discussed here, and 
it will certainly not by itself explain such outstanding mysteries 
as the high abundances of CH\p, NH, \hcop, {\it etc.}
For the former, the reaction C + \H3\p $\rightarrow$ CH\p\ + \HH\ 
is exothermic but not particularly fast and the fraction of 
gas-phase carbon which is neutral is very small in diffuse gas,
typically below 1\%: this forms only  miniscule amounts of 
CH\p\ (more CH\p\ might be formed by the reaction of C with \HH\p).
For NH \citep{HawWil97,MeyRot91}, the reaction of N and \H3\p\ is 
endothermic.  For \hcop, which in dark clouds is formed {\it via}
the reaction \H3\p\ + CO $\rightarrow$ \hcop\ + \HH, the amounts
of both \H3\p\ and CO are too small in diffuse clouds to sustain
\hcop\ against its recombination to CO, 
\hcop\ + e $\rightarrow$ CO + \HH. CO can form from \hcop\ in 
quiescent diffuse gas \citep{LisLuc00}, but not the other way around.

One species for which \H3\p\ might make a difference is OH, because 
both the reactions O + \H3\p\ $\rightarrow$ OH\p\ + \HH\ (rate 
constant $k_1 = 8 \times 10^{-10}$ cm $^3$ \ps) and 
O\p\ + \HH\ $\rightarrow$ OH\p\ + H (rate constant
$k_2 = 1.7 \times 10^{-9}$ cm $^3$ \ps) are exothermic.  Figure 6 
shows the time evolution of the volume formation rate of OH\p\ through
both routes (shown are $k_1$n(O)n(\H3\p) and $k_2$n(O\p)n(\HH)) 
in three models threaded by a high cosmic ray flux, 
\zetaH\ $= 3 \times 10^{-16}$ \ps, at three densities 
n(H) = 32 $\pccc$, 64 $\pccc$ and 128 $\pccc$.  The cloud diameter 
is held fixed so that N(H) $= 4\times10^{20}~\pcc$, 
$8\times10^{20}~\pcc$ and $ 16\times10^{20}~\pcc$, respectively, 
purposely driving the highest density gas to very high \HH\ fractions 
and \H3\p\ abundances at late times.

As shown in Fig. 6, the reaction of O and \H3\p\ is never a dominant 
source of OH\p.  The figure does show, however, the effect of ionization
balance on the OH formation rate (see also Sect. 2.3 of \cite{Lis03}): 
at later times the first step toward OH formation occurs more vigorously 
in the more tenuous models because their higher ionization levels 
(both overall and in oxygen) more than compensate for lower \HH\ 
fractions.  In the models with larger \HH\ fractions, the proton density 
declines at late times because a more nearly molecular gas cannot 
sustain a high proton density and the O\p/O ratio also declines, 
due partly to its rapid charge exchange reaction with hydrogen 
and because of its rapid reaction with \HH.  The cosmic-ray ionization
alone, even at the high rates assumed here, cannot support high
ionization fractions in atomic hydrogen or oxygen in the presence of 
PAH and high \HH\ concentrations.

%6 was here

%6
\begin{figure}
\psfig{figure=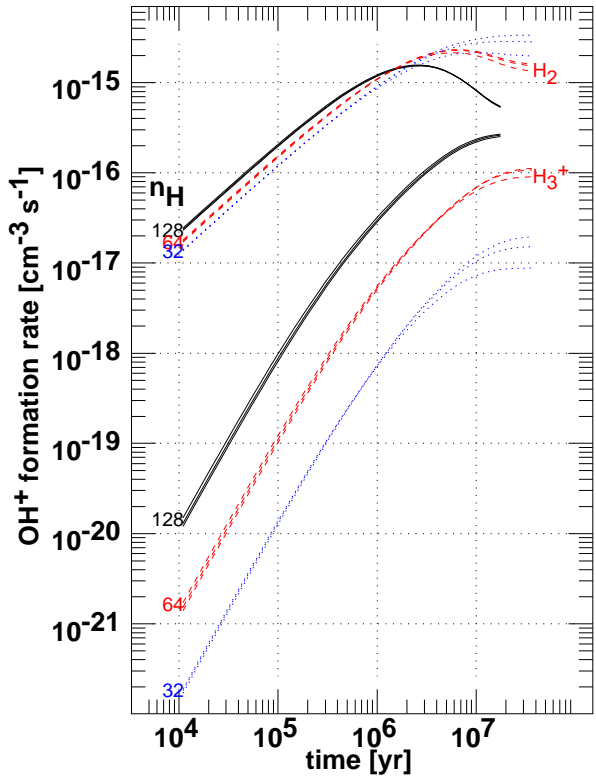,height=10.3cm}
\caption[]{OH$^+$ volume formation rates due to reaction of 
O\p\ + \HH\ (upper family of curves) and O + \H3\p\ (lower
family) in the presence of a high primary cosmic ray ionization 
rate \zetaH\ $= 3 \times 10^{-16}$ \ps.  Results are shown at 
the usual three fractional radii within models of diameter 4.05 pc 
having n(H) = 32 $\pccc$, 64 $\pccc$ and 128 $\pccc$ , so that
N(H) $= 4\times10^{20}~\pcc$, $8\times10^{20}~\pcc$
and $ 16\times10^{20}~\pcc$, respectively). }
\end{figure}

The next (necessary) step toward OH formation,  
OH\p\ + \HH $\rightarrow$ O\HH\p\ + H  will dominate over 
recombination of the OH\p\ ion for \HH-fractions as small as 2\%
at diffuse cloud temperatures.  Thus, the thinner models will 
actually form OH more rapidly (not just OH\p), as the \HH\ accumulation 
process equilibrates \footnote{At early times, the formation of OH in 
less dense clouds  will be seriously impeded by their larger e/\HH\ ratios 
and the denser models may make more OH.  As with \H3\p\ itself, high OH 
abundances only appear at the conclusion of the \HH\ accumulation process}.
Chemical effects like those shown in Fig. 6 undoubtedly contribute to 
the appearance of radio frequency OH emission in the envelopes of dark 
clouds well outside the most strongly molecular regions (see 
\cite{WanAnd+93} and Fig. 6 of \cite{LisLuc96}).  In the current context 
we expect/predict that regions of higher than normal OH abundance should 
be found optically in diffuse clouds having modest extinction and 
\HH\-fractions.  

\section{Summary and discussion}

\subsection{ Summary}

We showed the approach to equilibrium under benign conditions
for some recently-published models \citep{LisLuc00,Lis02,Lis03},
calculating the time-dependent \HH\ and \H3\p\ formation in pc-size,
otherwise largely static gas parcels meant to represent typical 
quiescent diffuse ``clouds''.   We pointed out that self-shielding 
does not hasten the formation or accumulation of \HH, although it may 
foster the production of {\it more} \HH\ over longer times.  Because
the \HH\ formation process in diffuse gas is incomplete, timescales 
can in general not be deduced from simple scaling arguments based on 
the local \HH\ formation rate.  The equilibrium \HH\ formation 
timescale in an H I cloud having a column density 
N(H) $\approx 4\times10^{20}~\pcc$ is $1-3\times10^7$ yr, nearly 
independent of the assumed density or \HH\ formation rate on grains, 
{\it etc}.  Under the benign conditions described here, the
bulk of the \HH\ formation in the diffuse ISM actually occurs 
under diffuse conditions in physically larger regions of moderate 
density; that is, there is little apparent advantage in hypothesizing 
a process which forms \HH\ in denser clumps, if the density remains
 within the range inferred for quiescent diffuse gas.

Other species reach equilibrium essentially as soon as \HH\ does, 
even when they involve higher-order and/or very ``slow'' processes
(like cosmic ray ionization of \HH) but their equilibrium abundances 
appear more nearly at just the last minute; it is more critical for 
explaining \H3\p\ that conditions equilibrate than, say, for \HH\ 
itself.  This is also relevant because the chemistry of a weakly-molecular
gas -- which could be either a model of low density or one of higher 
density at early time -- behaves in various particular ways.  For
instance, the predicted abundance of \H3\p\ is less strongly dependent 
on the cosmic-ray ionization rate when n(\HH)/n(H) $\la 0.05$.

Attempts to form increased amounts of \H3\p\ may be thwarted at very 
high assumed values of \zetaH\ both because the fractional ionization 
increases and because the \HH-fraction declines as cosmic-ray 
ionization overtakes photodissociation as the main \HH-destruction 
mechanism.  Assuming \zetaH $>> 10^{-16}\ps$ will also change the 
thermal balance in the diffuse ISM as a whole as cosmic-ray 
ionization becomes the dominant heating process in both warm and 
cold neutral diffuse gas. Although observations of both HD and 
\H3\p\ seem to indicate an enhanced low-level source of hydrogen 
ionization, it is important to ascertain whether it is indeed a 
cosmic-ray related process which is responsible (or some other
form of very hard radiation) and whether any related effects occur
 in other phases of the interstellar medium, either in the warm or 
dark gas.

The presence of relatively high \H3\p\ abundances {\it per se} probably 
does 
not have strong effects on the chemistry of other trace species whose 
abundances are puzzling (CH\p, NH, \hcop\ {\it etc.}) or on those 
of species like OH which are generally considered to be better 
understood.  However, in considering competition between OH formation 
paths through \HH\ and \H3\p, we showed that late-time OH formation 
proceeds most vigourously in more diffuse regions having modest 
density, extinction and \HH\ fraction and somewhat higher fractional 
ionization, suggesting that high OH abundances might be found 
optically in diffuse clouds having modest extinction.

\subsection{ Lifetimes of diffuse clouds and their \HH}

 In this work we modelled the formation and accumulation of \HH\
under conditions like those which are inferred from observations 
of the \HH\ and related neutral gas species themselves; temperatures 
from \HH\ \citep{SavDra+77,ShuAnd+04} and mean thermal pressures 
(densities) from C I \citep{JenTri01}.  Although modest amounts
of \HH\ form quickly enough that molecule-free sightlines are
relatively rare, the observed high mean molecular fractions of 
25\%-40\% \citep{SavDra+77,LisLuc02} are achieved only over
timescales of $10^7$ yr or longer, when starting from scratch.

\HH\ aside, timescales for thermal, ionization, chemical $etc.$
equilibrium are short \citep{Spi78,WolMcK+03,RawHar+02} and diffuse
gas is not normally thought of as having much longevity in any
particular form.  The H I cloud spectrum evolves under the continual
influences of cloud-cloud collisions, evaporation, passage of supernova 
blast waves and stellar wind shocks $etc.$ \citep{ChiLaz80} and in a 
3-phase model of the diffuse ISM, gas in any particular volume element 
turns over on timescales shorter than $10^6$ yr according to \cite{McKOst77}.

How then are high \HH\ fractions and short diffuse gas lifetimes to
be reconciled?  This question has never really been considered when 
the overall structure of the diffuse ISM is modelled, but there are 
several alternatives.  1) The long equilibrium timescales under diffuse 
neutral conditions provide, in reverse, some resilience which might 
help pre-existing \HH\ to weather the changes which occur.  2) Dark clouds 
have haloes \citep{WanAnd+93,Ben06} and it has been suggested that the 
molecules supposedly seen in diffuse clouds are actually located around 
and exchanging material with dark clouds \citep{FedAll91};  this may be 
true in some cases, but complex trace molecules are also seen along 
sightlines which are quite well separated from the nearest dark material 
\citep{LisLuc+06}.  3) Perhaps most promisingly, turnover within 
diffuse neutral gas due to turbulence, rapidly cycling material through 
a very dense phase, may provide for much faster formation of \HH\ and trace 
species \citep{GloMac05,FalPin+05}, establishing high molecular fractions
from scratch on short timescales.  The small fraction of neutral material 
which is observed to exist at very high pressure within diffuse clouds 
\citep{JenTri01} and other aspects of small-scale structure in diffuse
gas \citep{Des00} may perhaps also be understood in such terms.

\begin{acknowledgements}

The National Radio Astronomy Observatory is operated by Associated 
Universites, Inc. under a cooperative agreement with the US National 
Science Foundation.  The calculations described here were mostly completed 
while the author was enjoying the hospitality of Bel 'Esperance and 
La Clemence in Geneva and the Hotel de la Paix in Lausanne.  The author
is grateful to the referee and the editor for several stimulating comments.

\end{acknowledgements}
 
%\bibliographystyle{apj}
%\bibliography{mnemonic,absorption}

\end{document}